\documentclass[pra,twocolumn,showpacs,preprintnumbers,floatfix]{revtex4}
\usepackage{graphicx}
\usepackage{dcolumn}
\usepackage{bm}
\usepackage{latexsym,epsfig}
\usepackage{graphicx}
\usepackage{verbatim}
\usepackage{comment}
\usepackage{amsmath}
\usepackage{amssymb}
\usepackage{stmaryrd}
\usepackage{color}

\newcommand{\beq}{\begin{equation}}
\newcommand{\eeq}{\end{equation}}
\newcommand{\bea}{\begin{eqnarray}}
\newcommand{\eea}{\end{eqnarray}}
\newcommand{\ben}{\begin{eqnarray*}}
\newcommand{\een}{\end{eqnarray*}}
\newcommand{\bfig}{\begin{figure}}
\newcommand{\efig}{\end{figure}}

\begin{document}
\title{Polar molecules in frustrated triangular ladders}
\author{Tapan Mishra}
\affiliation{Institut f\"ur Theoretische Physik, Leibniz Universit\"at Hannover, 30167~Hannover, Germany}
\author {Sebastian Greschner}
\affiliation{Institut f\"ur Theoretische Physik, Leibniz Universit\"at Hannover, 30167~Hannover, Germany}
\author{Luis Santos}
\affiliation{Institut f\"ur Theoretische Physik, Leibniz Universit\"at Hannover, 30167~Hannover, Germany}

\date{\today}

\begin{abstract}
Polar molecules in geometrically frustrated lattices may result in a very rich landscape of quantum phases, due to the non-trivial interplay between 
frustration, and two- and possibly three-body inter-site interactions. In this paper, we illustrate this intriguing physics 
for the case of hard-core polar molecules in frustrated triangular ladders. Whereas commensurate lattice fillings result in gapped phases with bond-order and/or density-wave order, at incommensurate 
fillings we find chiral-, two-component-, and pair-superfluids. We show as well that, remarkably, polar molecules in frustrated lattices allow, for the first time to our knowledge,
for the observation of bond-ordered supersolids.
\end{abstract}


\maketitle

\section{Introduction}
Ultracold quantum gases in optical lattices constitute an excellent scenario for the study of strongly-correlated many-body systems~\cite{lewenstein_book}, as highlighted by 
the superfluid~(SF) to Mott-insulator transition of cold atoms in optical lattices~\cite{bloch} driven by the interplay between inter-site hopping  
and two-body onsite interactions. Two-body inter-site interactions are expected to play a crucial role in polar lattice gases, including magnetic  
atoms, polar molecules, and Rydberg atoms~\cite{pfaureview,ni,gallagher}. Inter-site dipole-dipole interactions may result under proper conditions in 
crystalline and supersolid phases~\cite{pfaureview,Baranov2012}. 

In typical experiments up to now multi-body interactions have played a negligible role compared to two-body ones. 
It has been however recently proposed that three-body nearest-neighbour interactions may be achieved under proper conditions in lattice gases of 
polar molecules~\cite{buchler}~(for a recent proposal on how to induce three- and even four-body on-site interactions in alkali lattice gases see Ref.~\cite{petrov2}). 
The seminal proposal of Ref.~\cite{buchler} opened fundamental questions about the effect of inter-site three-body interactions in lattice gases. In particular, recent works have shown that 
three-body interactions may result in supersolid and Devil's staircase phases in 2D~\cite{lauchli} and in a phase transition in 1D systems from a SF to a gapped phase with simultaneous 
charge-density-wave~(CDW) order and bond-order~(BO)~\cite{sansone}. 

Frustrated lattice systems constitute nowadays a very active research focus, due to recent experimental progresses in the creation of 
synthetic gauge fields~\cite{spielman,Aidelsburger2013,Miyake2013} and the possibility of realizing 
frustrated lattice geometries~\cite{Struck2011}. The rich physics resulting from the 
interplay between frustration and interactions has attracted since years a growing theoretical 
attention~\cite{giamarchi_meisnereffect,dhar1,dhar2,greschner,lehur,altman,paramekanti,dhar3,mishra_ttpvss,mishra_ttpv, piraud2014}. 
Recently, experiments on dynamically frustrated optical lattices in the presence of a synthetic gauge field have investigated the 
on-set of chirality, by observing the Meissner- to vortex-phase transition~\cite{bloch2014}.

In this paper, we are interested in the rich physics that results from the interplay between frustration and 
inter-site interactions in frustrated polar lattice gases. We illustrate the physics that may result from that interplay by 
considering the particular case of hard-core polar molecules in frustrated triangular ladders. By means of extensive numerical calculations 
based on density matrix renormalization group (DMRG) techniques~\cite{white_92,schollwock_review_05}, we show that the frustrated polar system 
is characterized by the appearance of commensurate gapped phases with bond-order and/or density-wave order, and by 
chiral-, two-component-, and pair-superfluids at incommensurate filling. Moreover, we show that, remarkably, polar molecules in frustrated lattices may 
allow for the first realization of bond-ordered supersolids.

The structure of the paper is as follows. In Sec.~\ref{sec:Model} we introduce the lattice model we employ to illustrate the interplay of frustration and 
two- and three-body interactions in polar lattice gases. Sections~\ref{sec:Gapped} and \ref{sec:Gapless} discuss in detail the gapped and gapless phases, respectively, 
found in the ground-state phase diagram. Section~\ref{sec:Dilute} discusses the dilute limit, whereas Sec.~\ref{sec:Interacting} discusses an effective quasiparticle 
model in the strongly-interacting regime. Finally, in Sec.~\ref{sec:Conclusions} we briefly summarize our conclusions.


\bfig[t]
  \centering
  \includegraphics*[width=0.45\textwidth,draft=false]{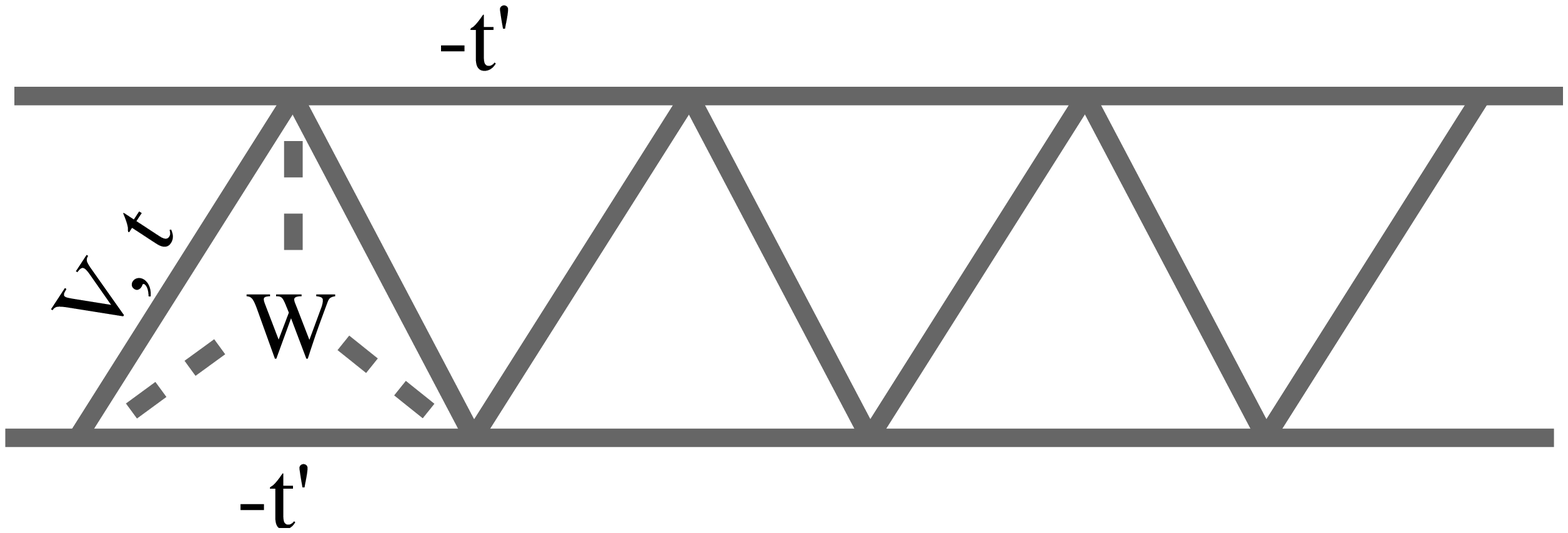}
    \caption{Triangular ladder lattice with two- and three-body interactions, $V$ and $W$, respectively. 
    The hopping along the rungs and the legs are $t>0$ and $t'<0$, respectively.}
    \label{fig:ladder}
\efig




\section{Model}
\label{sec:Model}

We consider in the following a system of polar molecules in a frustrated  triangular ladder as depicted in Fig.~\ref{fig:ladder}, 
characterized by hopping $t>0$ and $t'<0$ along the rungs and the legs, respectively. The change in sign of $t'$ may be achieved e.g. by 
lattice shaking as recently realized experimentally~\cite{Struck2011}. Such a ladder is actually equivalent to 
a 1D lattice with nearest-neighbor~(NN) hopping, $t$, and next-to-nearest-neighbor~(NNN) tunnelling, $t'$. As previously discussed, under proper conditions, 
polar molecules may present in addition to two-body inter-site interactions, which we restrict in the following to the NN interaction, $V$, 
significant three-body NN interactions, $W$, of molecules belonging to the same triangular plaquette (see Fig.~\ref{fig:ladder}).
In the following we explore in particular a simplified hard-core scenario, in which maximally one molecule may occupy each site. The latter condition 
assume large-enough on-site interactions and a lattice filling smaller than one.
The Hamiltonian describing the system is of the form:
\begin{eqnarray}
\!\!\!\!\!\! H \! &=& \!  - t \sum_i (a_{i}^{\dagger}a_{i+1}^{\phantom \dagger} + \text{H.c.})
 \! - \! t' \sum_i (a_{i}^{\dagger}a_{i+2}^{\phantom \dagger}+\text{H.c.}) \notag \\
 &+& V\sum_i  n_in_{i+1}+W\sum_i  n_in_{i+1}n_{i+2} 
\label{eq:ham}
\end{eqnarray}
where $a_i^{\dagger}$ and $a_i^{\phantom \dagger}$ are creation and annihilation operators
for bosons at site $i$, and $n_i=a_i^{\dagger}a_i^{\phantom \dagger}$ is the boson number operator at site $i$. 
We set below $t=1$ as the energy scale. 

Spin-systems with two-body interactions in such ladders
have been studied extensively both numerically and analytically~\cite{Kolezhuk, Vekua, Hikihara2000, Hikihara2002}, at half filling~\cite{Hikihara01, Furukawa10}, 
and with magnetic fields in the ferromagnetic~\cite{Hikihara08} as well as antiferromagnetic regime~\cite{Hikihara04, Hikihara10}.
Model~\eqref{eq:ham} presents no exact solution except for $W=t'=0$ for which there exists 
a SF to CDW phase transition at $V=2t$~\cite{rigol_giamarchi_rmp}, 
and for $W=V=0$ for which the ground state can be obtained exactly at $t'=-t/2$~\cite{mishra_ttpv}. 

In this paper we perform an extensive numerical study of model~(\ref{eq:ham}), and in particular 
of the interplay between frustration and two- and three-body interactions as a function of the lattice filling, $0<\rho<1$. 
We find a very rich landscape of ground-state phases, as illustrated in Fig.~\ref{fig:phasedia} for $t'=-1$ and $V=W$. 
For half filling, $\rho=1/2$, the competition between frustration and two-body interactions results in a phase transition   
between a BO phase and a CDW phase, which is of first-order nature, as predicted in Ref.~\cite{mishra_ttpv}. 
On the contrary, for $\rho=2/3$ the three-body interactions result in a Berezinskii-Kosterlitz-Thouless~(BKT) transition between a SF 
and a phase with both BO and CDW orders~(BO+CDW phase)~\cite{sansone}. 
In addition to these phases, we find a wealth of gapless phases at incommensurate fillings.
For small two- and three-body interactions, and all fillings(except for half filling), the systems becomes a chiral superfluid~(CSF) that carries a 
finite long-range current order~\cite{Hikihara10}. This current order vanishes with growing interactions, and the system enters into 
non-chiral SF phases, which include supersolid~(SS), 
two-component superfluid~(2SF), and pair-superfluid~(PSF) phases. 
Interestingly, these three phases exhibit coexisting non-vanishing peak in 
both BO and CDW structure factor. 
To the best of our knowledge, such bond-ordered supersolids (BO+SS) are reported here for the first time.
In the following section we discuss these phases in more detail.



\bfig[!t]
  \centering
  \includegraphics*[width=0.45\textwidth,draft=false]{phasedia1.eps}
    \caption{(Color online) Phase diagram of model~\eqref{eq:ham} as a function of $\mu'$ and $V=W$, for $t'=-t=-1$. Here $\mu'=\frac{9+4\mu}{20W+18}$ is the 
    scaled chemical potential such that $\mu'=0(1)$ corresponds to the vacuum~(full) state. 
    Solid (red) curves mark the boundaries of gapped phases, such as BO and CDW at $\rho=1/2$ and BO+CDW at $\rho=2/3$. 
    The CSF phase is present for all $\mu'$ (except for half filling) and weak interactions. The black circles separates the 
    CSF phase from other superfluid phases. The 2SF-SS transition is marked by green circles. 
    The shaded region corresponds to the cSS phase. 
    The PSF phase is a small region appearing in the upper part of the BO+CDW phase.}
    \label{fig:phasedia}
\efig




\section{Gapped phases}
\label{sec:Gapped}

We discuss in this and the next section in detail the phases depicted in Fig.~\ref{fig:phasedia} corresponding to the case $t'=-1$ and $V=W$. 
At half filling, $\rho=1/2$, the system is characterized by the appearance of gapped incompressible BO and CDW phases. As shown in Fig.~\ref{fig:phasedia}, 
these phases are separated by a first-order phase transition at $V=W\simeq 3.0$. 
The BO phase results from frustration that favours dimerization in the 
small $V$ regime, whereas large $V$ favours the CDW phase~\cite{mishra_ttpv}. Three-body interactions do not play any 
significant role in these phases since in them three molecules never coexist simultaneously at the same triangular plaquette.
The BO phase is characterized by a finite bond-order structure factor 
\begin{equation}
 S_{BO}(k)=\frac{1}{L^2}\sum_{i,j}{e^{ik(i-j)}\langle B_i B_j\rangle},
\label{eq:str}
\end{equation}
where $B_i=a_i^\dagger a_{i+1}^{\phantom \dagger}+a_{i+1}^\dagger a_i^{\phantom \dagger}$, whereas the CDW 
presents a finite value of the density-density structure factor 
\begin{equation}
 S_{CDW}(k)=\frac{1}{L^2}\sum_{i,j}{e^{ik(i-j)}(\langle{n_{i}n_{j}}\rangle -\langle n_i\rangle\langle n_j
\rangle)}.
\label{eq:bostr}
\end{equation}
at the wave vector $k=\pi$. At the transition point the excitation gap diminishes but never vanishes showing 
the first-order nature of the phase transition. 

For $\rho=2/3$ the system presents another gapped incompressible phase, BO+CDW, that  
possesses both BO and CDW orders simultaneously. In the absence of $t'$ and $V$ the system undergoes a SF-to-(BO+CDW) BKT transition at 
$W \sim 3.0$~\cite{sansone} where the Luttinger liquid parameter $K=2/9$, and $S_{CDW}(k)$ has a peak at $k=2\pi/3$. 
We observe that, in the presence of both frustration and inter-site two-body interactions the BO+CDW phase is enhanced, and 
hence the SF-(BO+CDW) transition occurs at a much smaller value, $V=W\simeq 1.0$~(see Fig.~\ref{fig:phasedia}). 
This happens because lattice frustration favours BO whereas two-body inter-site interactions favour CDW. 

In order to distinguish between gapped and gapless phases we evaluate the chemical potential
$\mu$ obtained from the minimization of $E(L,N)-\mu N$ where  
$E(L,N)$ is the ground-state energy of the system with $L$ sites and $N$ bosons~\cite{Hikihara08}.
In Fig.~\ref{fig:cuts}(a) we plot $\mu$ as 
a function of $\rho$ for $L=120$ and $V=W=2.6$. It can be seen that $\mu$ jumps at
$\rho=1/2$ and $\rho=2/3$, indicating the gapped phases. 



\bfig[htbp]
  \centering
  \includegraphics*[width=0.42\textwidth,draft=false]{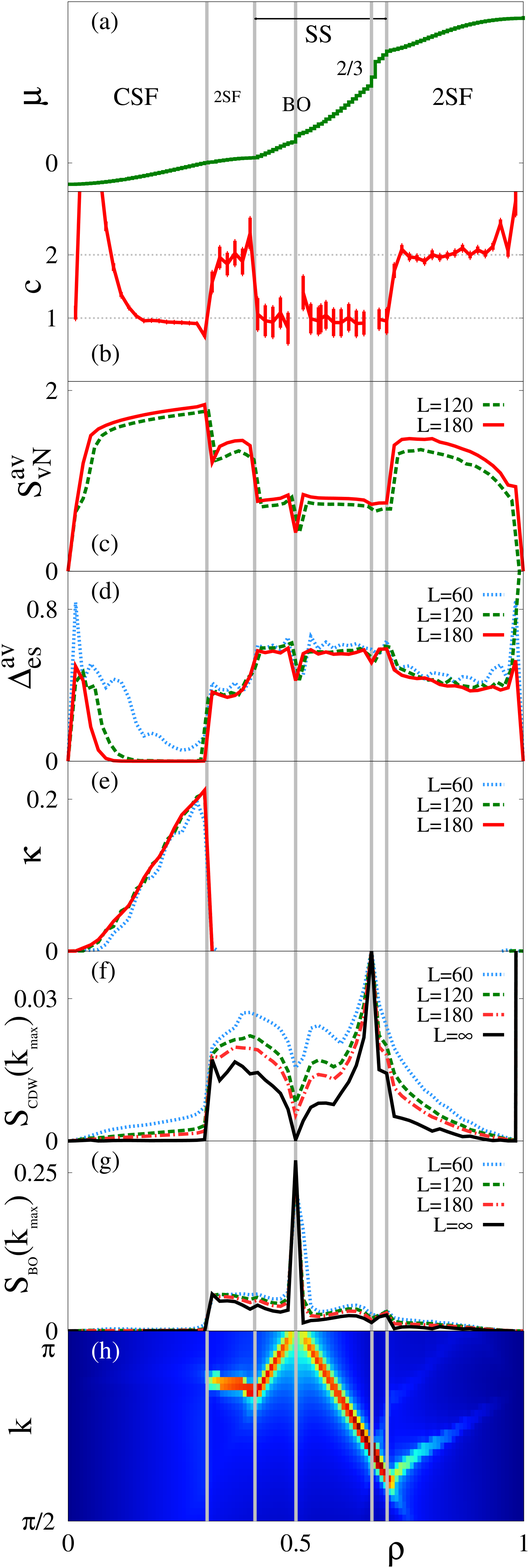}
    \caption{(Color online) Cut through the phase diagram of Fig.~\ref{fig:phasedia} for $V=W=2.6$. We depict as a function of $\rho$:  
    (a) Chemical potential $\mu$ ($L=120$), (b) central charge $c$ ($L=180$), (c) entanglement entropy, (d) entanglement gap, 
    (e) chirality $\kappa$, (f) maximum of $S_{CDW}(k)$, (g) maximum of $S_{BO}(k)$ (both (f) and (g) are obtained by means of a linear extrapolation to $L\to\infty$ 
    using the two largest system sizes $L=120$ and $L=180$). The panel (h) depicts $S_{CDW}(k)$ as function of (quasi)momentum $k$ and filling $\rho$ ($L=120$).}
    \label{fig:cuts}
\efig


\section{Gapless phases}
\label{sec:Gapless}

\subsection{2SF phases}

For $\rho\neq 1/2$ or $2/3$, the system is characterized by different types of SF phases. In order to understand in more detail the 
nature of these phases, we consider a fixed $V=W=2.6$ and plot various relevant physical 
quantities as a function of $\rho$ in Figs.~\ref{fig:cuts}. 
As discussed above, $\mu(\rho)$~(Fig.~\ref{fig:cuts}(a)) shows the appearance of 
incompressible (gapped) and compressible (gapless) phases. A first differentiation between gapless phases is provided by the study of the von Neumann entropy: 
\begin{equation}
S_{vN, L}(l) = -\mathrm{tr}\left( \rho_l \ln\rho_l \right) = \frac{c}{6} \ln\left[ \frac{L}{\pi} \sin\left(\frac{\pi}{L}l\right) \right] + g
\label{eq:ent} 
\end{equation}
where $\rho_l$ is the reduced density matrix for a subsystem of length $l$ embedded in a chain of a 
finite length $L$, and the last equation is valid for conformally invariant gapless states~\cite{vidal03, calabrese04}, with 
$c$ the central charge, and $g$ a constant. Figure~\ref{fig:cuts}(b) shows the value of $c$ extracted by the finite size scaling of $S_{vN, L}(l)$.
We observed the apperance of regions with $c=1$~(SF phases with a single gapless mode), and regions with $c=2$~(and hence two gapless modes). We denote the later 
as 2SF phases~\cite{footnote-c}. 

Apart from estimating $c$, both entanglement entropy $S_{vN,l}$ but also the spectrum of eigenvalues $s_i^2(l)$ of the reduced desity matrix $\rho_l$ of the subsystem itself - the so called entanglement spectrum - have been shown to offer a sensitive probe for quantum phase transitions~\cite{calabrese04, Li08, Pollmann10, Poilblanc10, Deng11, Chiara12}. 
We have computed in particular the gap in the entanglement spectrum, 
$\Delta_{es}(l) = \sum_i (-)^i s_i(l)$. In order to minimize boundary effects, we have averaged  
$S_{vN, L}(l)$ and $\Delta_{es}(l)$ over all bipartitions in the range $[L/4, 3L/4]$. We denote these averages as $S_{vN}^{av}$ and $\Delta_{es}^{av}$. 
The 2SF-SF transitions are marked by sharp jumps in $S_{vN}^{av}$ and $\Delta_{es}^{av}$ (see Figs.~\ref{fig:cuts}(c) and (d)). 
Interestingly, these transitions are also characterized by a kink in the $\mu(\rho)$ plot of Fig.~\ref{fig:cuts}(a).

\subsection{Chiral Superfluid Phases}

Geometric frustration dominates the system for weak interactions $V=W$. In this regime, the frustration leads to a CSF phase 
for all values of $\rho$~(except $\rho=1/2$). The CSF phase eventually vanishes for sufficiently large interactions~(Fig.~\ref{fig:phasedia}). 
CSF phases are characterized by a non-zero chiral order parameter:
\beq
\kappa=\lim_{\left|i-j\right|\gg1} \left<\kappa_i\kappa_j\right>
\label{eq:kappa}
\eeq
where $\kappa_i=i(a_i^\dagger a_{i+1}^{\phantom \dagger} -a_{i+1}^\dagger a_i^{\phantom \dagger})$. 
As shown in Fig.~\ref{fig:cuts}(e) $\kappa>0$ for low fillings (CSF phase), vanishes sharply at the onset of a 2SF phase, and remaining zero for all 
larger $\rho$ values. 
The CSF phase is characterized by the spontaneous symmetry breaking of the $Z_2$ symmetry associated to the two dispersion minima induced 
by the lattice frustration~\cite{greschner}. This spontaneous symmetry breaking nature of the CSF phase becomes apparent from the vanishing $\Delta_{es}^{av}$~(Fig.~\ref{fig:cuts}(d))
(note, however, that $\Delta_{es}^{av}$ shows strong finite size corrections especially at low $\rho$).

\subsection{Supersolid Phases / Bond-ordered supersolids}

Supersolids are characterised by the coexistence of CDW order and superfluidity. 
Supersolids can be achieved in bosonic lattice systems with nearest neighbour 
interactions~\cite{pfaureview,Baranov2012}. 
Whereas at commensurate densities the system undergoes for large-enough interactions a SF-CDW transition, a SS may be obtained by doping 
the system away from the crystalline phases. In 1D, hard-core bosons do not exhibit supersolidity due to large fluctuations and the hardcore repulsion. 
However, kinetic frustration can contribute significantly to stabilize SS phases~\cite{mishra_ttpvss}. 

In contrast to the case of soft-core bosons, the SS in the hard-core case are characterized by a peak in $S_{CDW}$ at a value $k_{max}$ incommensurate to the lattice, 
and dependent on the filling $\rho$~\cite{rossini,mishra_ttpv}. As mentioned above, in the present case, there exists two CDW phases with $S_{CDW}$ peaked at $k_{max}=\pi$~($\rho=1/2$) 
and $2\pi/3$~($\rho=2/3$). By doping the system away from these CDW phases we retain a finite CDW order in the adjacent SF region that interpolate between the $k_{max}$ values in the CDW regions. 
Figure~\ref{fig:cuts}(h) shows the linear $\rho$ dependence of $k_{max}$ in the superfluid regions adjacent to the CDW phases.
The presence of crystalline order is confirmed by a finite value of $S_{CDW}(k)$ in the thermodynamic limit~(see Fig.~\ref{fig:cuts}(f)). 
It can be seen that $S_{CDW}(k)=0$ in both the CSF phase and the BO phase at $\rho=1/2$.
The presence of CDW order in the superfluid phases confirms the existence of supersolidity~(note that 2SF has a finite CDW order, and hence it is also a supersolid phase).

Remarkably, in addition to a finite CDW order we observe a finite coexisting bond order.
Figure~\ref{fig:cuts}(g) depicts the maximum of $S_{BO}(k)$ as a function of $\rho$.   
Clearly the system exhibits simultaneous CDW and BO orders not only at the gapped BO+CDW at $\rho=2/3$, but for incommensurate fillings within the superfluid region 
as well ~(except in the CSF region where both BO and CDW orders vanish).
We would like to mention that the bond ordering in the system of hardcore
bosons arises due to the effect of frustrated next-nearest neighbour hopping $t'$ at low
filling around $\rho=1/2$. However, in the high filling regime, the bond ordering occurs due to the effect of the three-body interaction $W$.
It can be recalled that in the absence of $t'$ and $V$, the system undergoes a SF-(BO+CDW) phase transition at $\rho=2/3$
filling~\cite{sansone}. The presence of additional $V$ only renormalizes the critical point of this transition.
As we depart from these commensurate densities, the system exhibits bond ordering at large frustration and interaction.
The BO+SS phase is the result of both frustration and interaction in the model considered here.
To the best of our knowledge, this constitutes the first instance of a BO+SS phase ever found.



\begin{figure}[t]
\includegraphics[width=0.8\columnwidth]{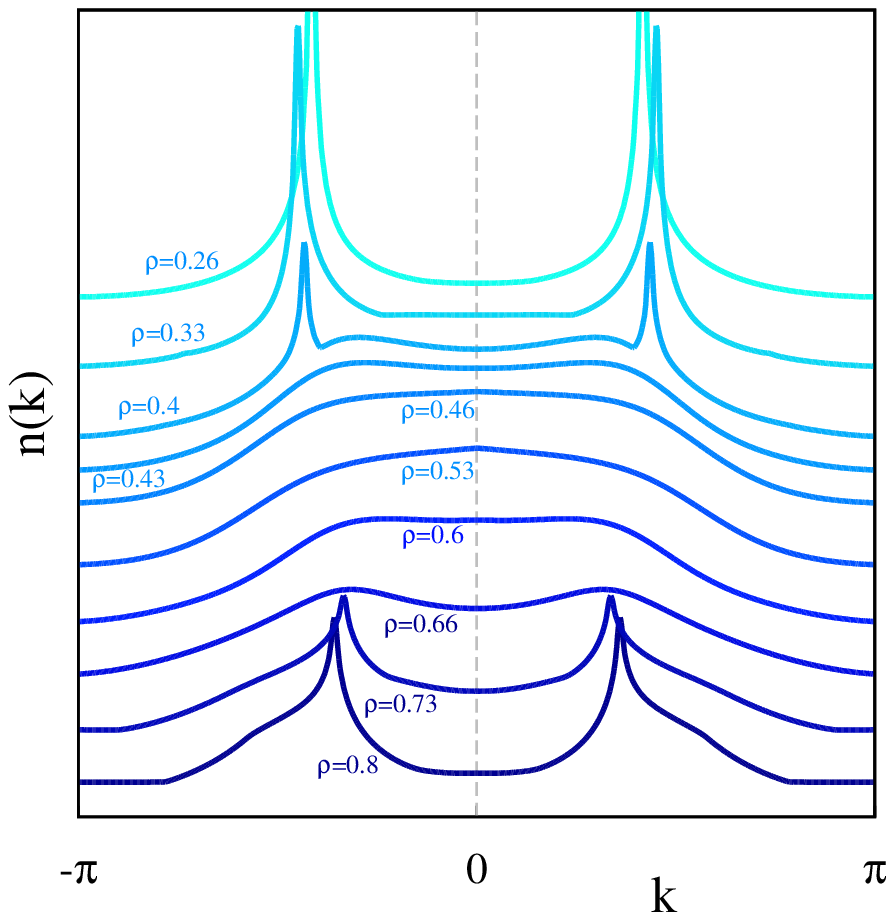}
\caption{Momentum distribution $n(k)$ for different densities ($V=W=2.6$, $L=180$). 
The single curves have been shifted for clarity. In the 2SF ($\rho=0.33$, $0.4$, $0.73$, and $0.8$) and the CSF phase ($\rho=0.26$), $n(k)$ 
exhibits two peaks at incommensurate momenta $\pm k$. When approaching the SS phase the momentum distribution broadens but 
still shows two maxima~(for $\rho=0.4$, $0.43$, $0.6$, and $0.66$), characterizing the icSS region. However, at intermediate fillings~($\rho=0.53$) only one maximum is observed (cSS region).}
\label{fig:mom_L180_V2.6_p3d}
\end{figure}


\subsection{Commensurate and incommensurate supersolids}

Another interesting feature of the supersolid phases is revealed by the momentum distribution:
\beq
n(k)=\frac{1}{L}\sum_{i,j}{e^{ik(i-j)}\langle{a^{\dagger}_{i}a_{j}}\rangle}.
\label{eq:mom}
\eeq
Fig.~\ref{fig:mom_L180_V2.6_p3d} depicts $n(k)$ for different $\rho$ for the same case of Figs.~\ref{fig:cuts}.
While in the 2SF and CSF region we observe two peaks at 
incommensurate momenta $\pm k$~\cite{greschner}, 
in the SS region $n(k)$ significantly broadens. 
Interestingly we find a wide SS region that still 
exhibits two maxima in $n(k)$, while in other regions of the 
SS phase $n(k)$ presents a single maximum at $k=0$. 
We call the former incommensurate supersolid~(icSS) and the latter commensurate supersolid~(cSS). The cSS region~(shaded 
area in Fig.~\ref{fig:phasedia}) grows slowly with increasing $V=W$~(e.g. we do not observe a single maximum for $\rho>2/3$ for $W=V\lesssim 20$). 
However, as we show below, in the limit of strong interactions only the cSS phase survives.

\subsection{Pair-superfluid phase}

Apart from the CSF and SS phases we observe a small pair-superfluid~(PSF) region in the phase diagram of Fig.~\ref{fig:phasedia}.
PSF phases are characterized by algebraically decreasing pair-pair correlations that coexists with exponentially decreasing single-particle correlations.
We would like to mention that the PSF region gets enhanced as $|t'|$ increases. As shown below, this may be also
well understood in the limit of strong interactions.



\begin{figure}[t]
\includegraphics[width=0.9\columnwidth]{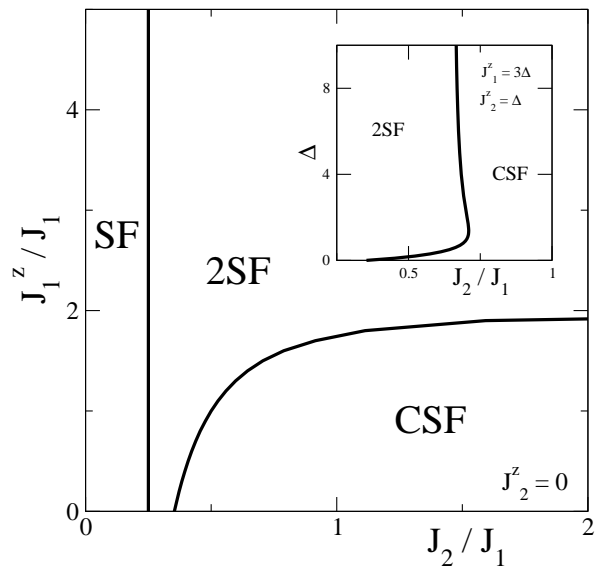}
\caption{Analytical prediction of the phase diagram for filling $\rho\to 0$ as function of the frustration $J_2/J_1$ and the interaction $\Delta=V=W$. 
The inset shows the phase boundaries for $J^z_1=3\Delta$ and $J^z_2=\Delta$ which corresponds to the particular case of $\rho\to1$ of model (\ref{eq:ham}). 
}
\label{fig:PD_dilute_BH_V1V2}
\end{figure}


\section{Dilute limit}
\label{sec:Dilute}

The CSF to 2SF transition is best understood in the dilute limit, $\rho\to 0$~\cite{Kolezhuk11}.
In this limit the single-particle dispersion in the triangular ladder develops two degenerate minima 
for $|t'/t|>1/4$. While for a single-minimum dispersion the system is a one-component SF, 
in the presence of two minima two different types of SFs may occur: either the bosons equally occupy both minima, 
which corresponds to a 2SF, or one of them is spontaneously selected. The latter is the case of the CSF, which hence 
exhibits a non-zero average momentum, a non-vanishing chiral current, and a sharp single peak in $n(k)$ away from $k=0$ in the thermodynamic limit~\cite{greschner}.

As shown in~\cite{Kolezhuk11} one may obtain quantitative insight in the competition between CSF and 2SF phases in the dilute limit from solving the low energy scattering problem of two bosons on the triangular ladder. From the two particle problem one may obtain different types of scattering solutions and extract two 
relevant scattering lengths: one for bosons belonging to the same single-particle minimum, $a_{1,1}=a_{2,2}$, 
and other one for bosons belonging to different 
minima $a_{1,2}$. One can relate the 1D scattering length to the amplitude of the 
contact interaction potential of the two-component Bose gas of mass $m$ as $g_{i,j}=-2/a_{i,j}m$. 
As the intra-component interaction is stronger than the inter-component 
interaction $g_{1,1}=g_{2,2}>g_{1,2}$ in the limit of vanishing density $\rho\to 0$ the 
 2SF phase is favoured. For a dominant inter-component 
interaction the particles would preferably occupy only one of the two minima that 
is spontaneously chosen, and hence the system would be in the CSF phase.

The quantitative analysis may be performed closely along the lines of Ref.~\cite{Kolezhuk11}. 
We may map the dilute bosonic model into a spin-$1/2$ model ($0,1 \rightarrow \uparrow,\downarrow$). 
The Hamiltonian then becomes 
a $J_1$--$J_2$ 
model with NN and NNN $S^zS^z$-interactions:
\begin{align}
H_{1/2}^{\rm dilute} &= J_1/2 \sum_i S^+_i S^-_{i+1} +  J_2/2 \sum_i S^+_i S^-_{i+2} + \text{H.c.}\nonumber\\
&+ J^z_1 \sum_i S^z_i S^z_{i+1} + J^z_2 \sum_i S^z_i S^z_{i+2} 
\label{eq:dilutej1j2spin12}
\end{align}
where $S_i^{x,y,z}$ denote the spin operators associated to the site $i$. 
For the low-filling limit, the spin couplings are given by $J^z_1=V=W$ 
and $J^z_2=0$, whereas $J_1/2=t$ and $J_2/2=t'$. 
The resulting phase diagram is shown in the main panel of Fig. \ref{fig:PD_dilute_BH_V1V2}. 
By increasing $J^z_1$ one observes the transition between the CSF and the 2SF phase for $J_2/J_1>1/\sqrt{8}$. 
The estimated value $J^z_{1,c} / J_1 \approx 1.8$, i.e. $V/t \approx 3.6$, for $J_2=J_1$ is consistent with our numerical 
estimates of the transition taken at finite density. 

Note that at high fillings just below saturation, $\rho\to 1$, we may again describe the system with the model~(\ref{eq:dilutej1j2spin12}).  
However, due to the broken particle-hole symmetry of model~(\ref{eq:ham}), the effective spin model presents different spin couplings,  
$J^z_1=3 \Delta$ and $J^z_2=\Delta$ with $\Delta=V=W$. 
As shown in the inset of Fig. \ref{fig:PD_dilute_BH_V1V2}, the CSF to 2SF transition at large fillings would be predicted for much lower values of $J_2/J_1$. 
However, in our numerical calculation we still find such a transition at $t=t'$ for high filling, 
which corresponds to the discrepancy already reported in Ref.~\cite{Kolezhuk11} for 
spin-$\frac{1}{2}$-systems with large values of $J^z_1$ and $J^z_2$.



\begin{figure}[t]
\includegraphics[width=0.9\columnwidth]{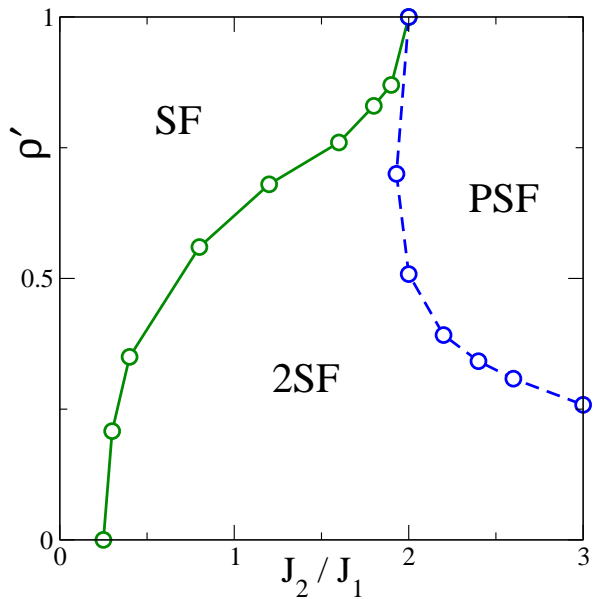}
\caption{(Color online)Phase diagram of model (\ref{eq:cHj1j2}) as a function of $J_2/J_1$ and the effective density $\rho'$.}
\label{fig:pd_cHj1j2}
\end{figure}


	

\bfig[h!tb]
  \centering
  \includegraphics*[width=0.42\textwidth,draft=false]{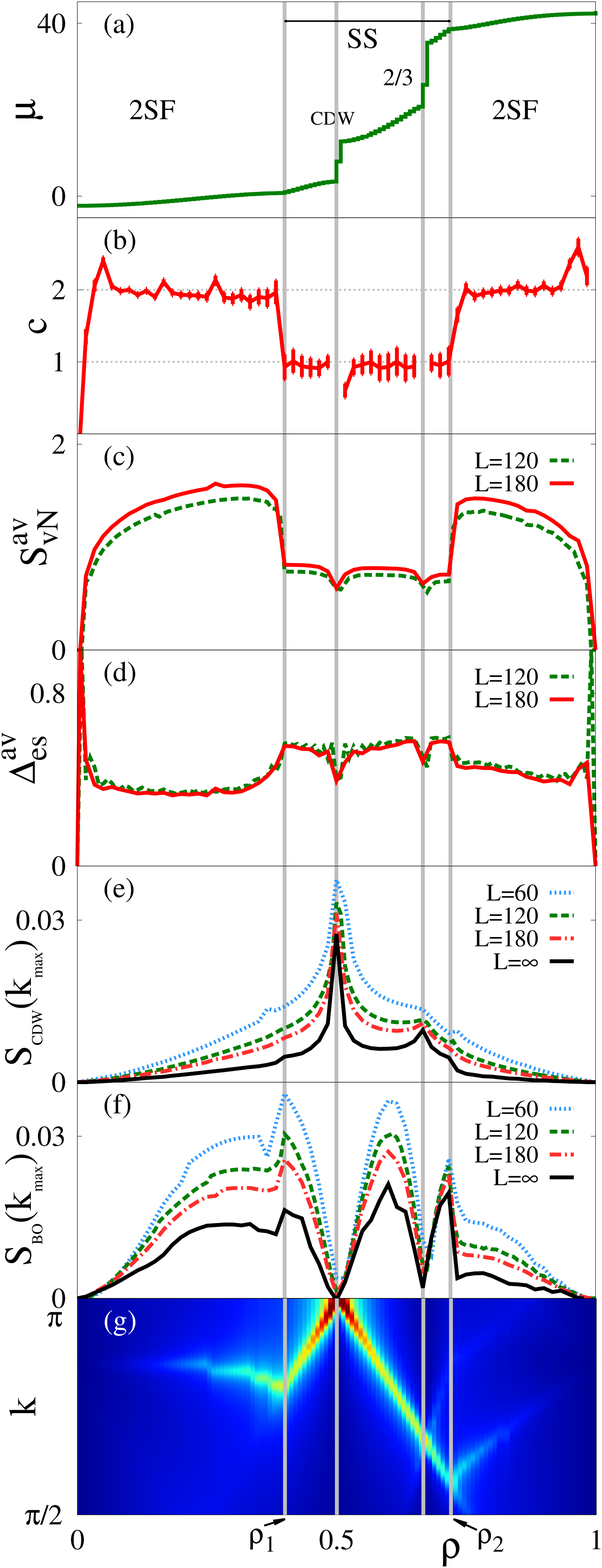}
    \caption{(Color online) Cut through the phase diagram of Fig.~\ref{fig:phasedia} at large interactions, $V=W=8.0$, and $|t'|=t=1$. We depict as a function of $\rho$: 
    (a) Chemical potential $\mu$ ($L=120$), (b) central charge $c$ ($L=180$), (c) entanglement entropy, (d) entanglement gap, 
    (e) maximum of $S_{CDW}(k)$, and (f) maximum of $S_{BO}(k)$. In panel (g) we show the structure-factor $S_{CDW}(k)$ as a function of (quasi)momentum $k$ and $\rho$ ($L=120$).}
    \label{fig:cuts_v8}
\efig


\section{Limit of strong interactions}
\label{sec:Interacting}

An interesting insight in the properties of the system is provided by the analysis of the strongly-interacting regime 
$V=W\gg t, t'$~(still assuming hardcore bosons). 
Let us first consider the case of low filling $\rho<1/2$, where three-body-interactions play a negligible role. 
In that regime we may identify pairs of subsequent $\left|01\right>$ particles as $\left| 1 \right>_{eff}$ and the 
remaining empty sites $\left|0\right>$ as $\left|0\right>_{eff}$. 
In this mapping the number of sites of the effective hardcore bosons is reduced to $L' = L-N$ 
and the total density of the effective model $\rho'$ fulfills $\rho=\frac{1}{1/\rho'+1}$. 
To first order in $(t,t')/V$ the effective quasiparticle model is given by an interaction-free $J_1-J_2$ 
Hamiltonian~(similar models have been studied on square lattices in Refs.~\cite{mila1,mila2}): 
\begin{equation}
H_{eff} = J_1 \sum_i c^\dagger_i c_{i+1} + J_2 \sum_i c^\dagger_i (1-c^\dagger_{i+1} c_{i+1}) c_{i+2} + \text{H.c.},
\label{eq:cHj1j2}
\end{equation}
where $c^\dagger$~($c$) are creation~(annihillation) operators for the effective quasi-particles. 
The correlated NNN tunnelling in Eq.~\eqref{eq:cHj1j2} stems from the fact that a hopping to 
site $i+2$ is only allowed if there is no neighbouring quasiparticle $\left| 1 \right>_{eff}$ on site $i+1$. 
Model~\eqref{eq:cHj1j2} also applies in the high-filling regime, $\rho>2/3$. In that case, we may identify tuples of three sites 
$\left|011\right>\to \left| 1 \right>_{eff}$ and the remaining occupied sites $\left| 1 \right>$
as $\left| 0 \right>_{eff}$. The effective length reduces to $L' = L-2N$, and the density is mapped as $\rho=1-\frac{1}{1/\rho'+2}$.
The phase diagram of model~(\ref{eq:cHj1j2}), depicted in Fig.~\ref{fig:pd_cHj1j2}, shows a SF-2SF transition for a critical $J_2/J_1<2$. 
We do not find any region with finite chirality. The SF phase of model(\ref{eq:cHj1j2}) may be identified with the SS phase of the original model due to the structure of the effective quasiparticles.

In Fig. \ref{fig:cuts_v8} we depict our numerical results obtained for the original model~\eqref{eq:ham} in the regime of large interactions ($V=W=8$ and $|t'|=t=1$). 
These numerical results agree well with those obtained from the strongly-interacting quasiparticle model~(\ref{eq:cHj1j2}).
For low and large fillings~($\rho<1/2$ and $\rho>3/2$) we observe a 2SF to SF transition, clearly revealed by the central charge $c$ and the entanglement 
properties~($S_{Vn}^{av}$ and $\Delta_{es}^{av}$).
The critical density for the SF-2SF transition obtained from the effective model matches well with the numerical estimates resulting from model~\eqref{eq:ham}. 
Figures~\ref{fig:cuts_v8} show SF-2SF transitions at $\rho_1\approx0.4$ and $\rho_2\approx0.72$ which agree well with $\rho_c'\approx 0.64$. 
For all incommensurate densities $S_{CDW}(k)$ and $S_{BO}(k)$ exhibit a maximum 
at $k\neq0$ which extrapolate to finite values in 
the thermodynamic limit~(the extrapolation has been performed with polynomials of first and second order in $1/L$).

As shown in Fig.~\ref{fig:pd_cHj1j2}, for large values of the NNN hopping $J_2$ it is energetically favorable at 
high densities to create pairs of holes~(PSF phase), because the correlated NNN hopping of 
 isolated quasi-holes is suppressed at high densities. This PSF phase at large NNN-hoppings is connected to the small 
 PSF region shown in Fig.~\ref{fig:phasedia}, and will occupy larger regions of the phase diagram with increasing $t'$. 
 A similar situation has been recently studied for the case of low fillings in strongly interacting dipolar lattice gases~\cite{Weimer14}.

For the particular choice of interactions $V=W$ the region of intermediate fillings $1/2<\rho<2/3$ in the large 
interaction limit may be mapped to a simple model of non interacting hardcore particles.  
By adding single particles on top of the perfect $\rho=1/2$ CDW phase one creates two 
domain-wall excitations that behave again as particle-pairs $\left|11\right>$ of two sites. However, this pair may only 
move by single sites with some amplitude $J$. It is precisely the large three-body interaction $W$ that
creates an effective hard-core repulsion of these excitations. Analogously one can start the description on the background of the perfect $2/3$-crystalline phase. 
Hence, the large interaction limit of the intermediate fillings is described by a simple non-interacting spinless fermion model on a chain, explaining 
why in the large interaction limit all SS regions map to a one-component standard SF phase. This phase exhibits a single 
maximum in the momentum distribution, broadened due to the size of the effective quasi-particles and thus at large interactions only the cSS phase is present.

\section{Conclusions}
\label{sec:Conclusions}

Polar molecules in geometrically frustrated lattices present a rich physics stemming from the interplay between frustration, and two-body and possibly three-body inter-site interactions. 
We have illustrated this physics for the particular case of hard-core polar molecules in frustrated triangular ladders. In addition to gapped phases $\rho=1/2$ and $\rho=2/3$ filling, 
we have revealed the appearance of a wealth of incommensurate superfluid phases, including chiral superfluids, two-component superfluids, pair-superfluids, and commensurate and incommensurate supersolids. 
Moreover, we have shown that except for the chiral superfluid phase, all superfluid phases exhibit coexisting bond and density-wave orders. To the best of our knowledge, this would be the first system 
realizing such bond-ordered supersolids.

Although we have just considered the specific case of model~(\ref{eq:ham}), our findings point to general features for models of polar molecules in the presence of geometric frustration, in particular the competition of 
CSF and 2SF phases, and the appearance of bond-ordered supersolids due to the simultaneous effect of frustration and inter-site interactions. Note in particular, that 
three-body interactions play a negligible role at low fillings $\rho<1/2$, such that the lower part of the phase diagram of Fig.~\ref{fig:phasedia} resembles to that of the 
$t-t'-V$-model~\cite{mishra_ttpv,mishra_ttpvss}. 

Finally, note that the predicted phases may be detected with state of the art techniques in optical lattice experiments. 
The chiral and non-chiral phases may be distinguished by vanishing and appearing peaks in time-of-flight images~\cite{Struck2011}. 
Kinks in the wedding-cake structure~(in the presence of an overall harmonic confinement) may be employed to monitor phase transitions into 
SS phases, which may be also revealed by a broadening of the TOF-peaks. Plateaus in the wedding-cake profile would show the commensurate-incommensurate 
transitions between gapless and gapped phases.

\acknowledgments
We acknowledge support by Center for Quantum Engineering and Space Time Research and by the DFG Research Training Group 1729. 
Computer simulations were carried out on the cluster system of the Leibniz Universit\"at Hannover.


\begin{thebibliography}{19}
\bibitem{footnote-c} The fit of Eq.~\eqref{eq:ent} to the block-entanglement data of finite systems with open boundary conditions 
is typically not very precise. However, we find that the fits to the numerical data 
for systems of length $L=180$ quite accurately yield the values of the central charge
 $c=1$ in the SF phase and $c=2$ in the 2SF phases, as shown in Fig. \ref{fig:cuts}(b). 
For low fillings we observe strong finite size corrections and the estimated values of $c$ apparently diverge.
\end{thebibliography}
\end{document}